# Light meson decay constants beyond the quenched approximation

G. M. de Divitiis, R. Frezzotti, M. Guagnelli, M. Masetti
and R. Petronzio

Dipartimento di Fisica, Università di Roma *Tor Vergata*

and

INFN, Sezione di Roma II

Viale della Ricerca Scientifica, 00133 Roma, Italy

October 25, 1995

## Abstract

We calculate the effects of including dynamical fermion loops in the lattice QCD estimates of meson decay constants, by extrapolating the results from negative flavour numbers after a suitable matching of the pion and rho mass. For moderately light quarks, the values of the decay constants not corrected for the renormalization constants increase with respect to their quenched values.

The computation of meson decay constants with lattice QCD simulations provides for light mesons a crucial test of the lattice approach and for heavy mesons the essential information which allows to extract the values of the still poorly known weak mixing angles.

While the estimates within the quenched approximation where internal fermion loops are neglected have reached a considerable degree of accuracy, those based on full QCD simulations are still affected by large errors which at present make the evaluation of the impact of the sea quark contribution difficult to extract.

In both cases, the extraction of the physical values of the decay constants requires the evaluation of operator renormalization constants which are known in most cases only perturbatively and the extrapolations of the final result to the continuum and, for light mesons, to the chiral limit. A recent analysis in the quenched approximation suggests [1] that the quenched results do not extrapolate to the correct experimental values, but lie somewhat below them.

In this letter we present a study of meson decay constants in the unquenched case with a method which allows to reach more precise estimates and a first evidence for sea quark effects.

The method, already discussed in refs. [2] and [3], extrapolates the unquenched results from a theory with a negative number of flavours to the physical case. The fields appearing in such a theory, called *bermions*, obey Bose statistics and are governed by the square of the hermitian operator $Q$, defined as $\gamma_5$ times the lattice Euclidean Dirac operator:

$$S[U, \phi] = S_g[U] + \sum_x \phi^\dagger(x) Q^2 \phi(x) \qquad (1)$$

where $S_g$ is the standard Wilson action for the gauge sector and $\phi(x)$ is the bermion field. Bermion fields have various indices which are omitted to shorten the notation: spin, colour and bermion–flavour. A theory with a positive number $n_b$ of bermion flavours corresponds to a negative number $n_f = -2n_b$ of fermion flavours. For the lattice Dirac operator we follow the standard Wilson formulation:

$$[Q\phi](x) = \frac{1}{2K}\gamma_5 \phi(x) - \frac{1}{2}\gamma_5 \sum_{\mu=0}^{3} U_\mu(x)(1-\gamma_\mu)\phi(x+\mu)$$
$$+ \frac{1}{2}\gamma_5 \sum_{\mu=0}^{3} U_\mu^\dagger(x-\mu)(1+\gamma_\mu)\phi(x-\mu) \qquad (2)$$

where $K$ is the usual Wilson hopping parameter related to the bare mass.

Such a theory exists in the Euclidean space only. The integration over bermion fields leads to a negative power of the determinant of the Dirac operator and therefore gauge configurations leading to small eigenvalues are sampled in the Monte

Carlo more frequently than in the fermion case where they are suppressed. In the limit of very light quark masses one can expect difficulties[1]. An example is given by the mass $m_S$ of the flavour singlet pseudoscalar. According to the Witten–Veneziano expression [4] $m_S^2 = m_{NS}^2 + (n_f/N_c)\lambda_\eta^2$, its mass for negative flavour numbers is lighter than the pion mass $m_{NS}$ and may lead to severe problems when approaching the chiral limit. We have monitored such a mass, indeed smaller than the pion mass, and in the results that we discuss we have made a safe choice of bare parameters. Detailed results on the pseudoscalar singlet will be presented elsewhere[5].

Changing the flavour content of the theory does change its lattice cutoff and, at fixed value of the Wilson hopping parameter $K$, the correction to the quark mass. The extrapolation from negative flavour numbers may be complicated by these variations. In order to minimize such an effect we have extrapolated theories after matching the values of two independent physical quantities, chosen to be the pion and the rho mass. The results that we present are obtained at two different values of the ratio $R2$ of the pion over rho mass squared, 0.7 and 0.5, and of the rho mass, 0.71 and 0.66, respectively.

The simulations were performed on a 25 Gigaflop machine of the APE series. The update procedure was for the gauge sector a Cabibbo–Marinari pseudo–heat [6] bath followed by three overrelaxation sweeps and for the bermion sector a heat bath followed by a number from three to seven overrelaxation sweeps[7].

The pseudoscalar and vector decay constants are defined in the continuum by

$$\langle 0|(\bar\psi_1\gamma_\mu\gamma_5\psi_2)_{cont}|\pi(p)\rangle = f_\pi p_\mu,$$
$$\langle 0|(\bar\psi_1\gamma_\mu\psi_2)_{cont}|\rho(p,\epsilon)\rangle = \epsilon_\mu \frac{m_\rho^2}{f_\rho} \quad (3)$$

where 1 and 2 are flavour indices and $\epsilon_\mu$ is the $\rho$ polarization vector. The operators chosen to extract the rho and pion decay constants are the standard lattice local operators:

$$\begin{aligned} P_5(\vec x, t) &= i\bar\psi_1(\vec x, t)\gamma_5\psi_2(\vec x, t), \\ V_k(\vec x, t) &= \bar\psi_1(\vec x, t)\gamma_k\psi_2(\vec x, t) \quad (k=1,2,3), \\ A_0(\vec x, t) &= \bar\psi_1(\vec x, t)\gamma_0\gamma_5\psi_2(\vec x, t), \end{aligned} \quad (4)$$

and the the decay constants were extracted from the behaviour at large times of the space integrated correlations given by

$$G_{55}(t) = \sum_{\vec{x}} \langle P_5(\vec{x},t) P_5^\dagger(\vec{0},0) \rangle,$$

$$G_{05}(t) = \sum_{\vec{x}} \langle A_0(\vec{x},t) P_5^\dagger(\vec{0},0) \rangle, \qquad (5)$$

$$G_\rho(t) = \sum_{k=1}^{3} \sum_{\vec{x}} \langle V_k(\vec{x},t) V_k^\dagger(\vec{0},0) \rangle$$

using the following formulae:

$$\frac{G_{05}(t)}{\sqrt{m_\pi G_{55}(t)}} \frac{\sqrt{e^{m_\pi T/2} \cosh(m_\pi(T/2-t))}}{\sinh(m_\pi(T/2-t))} \xrightarrow{t\ large} \frac{af_\pi}{Z_A}, \qquad (6)$$

$$\left[\frac{1}{3m_\rho^3} \frac{G_\rho(t)}{e^{-m_\rho T/2} \cosh(m_\rho(T/2-t))}\right]^{1/2} \xrightarrow{t\ large} \frac{1}{f_\rho Z_V} \qquad (7)$$

where $Z_A$ and $Z_V$ are the proper renormalization constants for the $A_0$ and $V_k$ local currents defined in eq. 5. The alternative correlation obtained by exchanging $A_0$ and $P_5$ in $G_{05}(t)$ definition gives the same average values but is more noisy. The quark propagator used to compute the hadron correlators was inverted every 1000 sweeps.

---

[1]This problem is not present if the bermions are subject to an external background gauge field like in the case of the Schrödinger functional. We thank M. Lüscher for this remark.

The modification of the renormalization constants due to fermion loops are expected to be a small correction. In the *leading log* approximation, the inclusion of fermion loops can be reabsorbed in a redefinition of the scale at which the renormalization constants are evaluated. Our matching procedure compensates for most of this scale change and leads to a very small residual $n_f$ dependence of these constants. In this paper, we give results not corrected for the renormalization constants which can be extrapolated in the flavour number without assumptions on their values. The fermion results can then be corrected with an appropriate fermion calculation.

We report in figure 1 and 2 the dependence of $f_\pi/(m_\rho Z_A)$ and $1/(f_\rho Z_V)$ upon the flavour number for $R2 = 0.7$ and $R2 = 0.5$ respectively, on a $16^3 \times 32$ lattice in the first case and also on a $24^3 \times 32$ lattice in the second one. The unquenched points and their errors are obtained from a linear fit with $\chi^2$ values ranging from 0.01 to 0.58. In Table 1 we report the same results. The corresponding values of the bare parameters used in the simulations of the different theories to obtain the matching, the number of configurations used to invert the Dirac operator and the values of rho and pion masses are given in Table 2. The values of the meson decay constants are extracted by a two steps fit where we first determine the meson mass from the large time behaviour of the correlation and then, by fixing the mass parameter in eqs. 6 and 7 to the best value, we determine the decay constant. We have checked that, by making a global four parameters fit which includes the excited states, the values of the mass and of the decay constant of the lightest state do not change beyond the statistical errors we quote.

The extrapolated value for $f_\pi/(m_\rho Z_A)$ at $R2 = 0.7$ can be compared with the one obtained from the direct unquenched fermion simulation described in ref. [8], at nearly matched values of the pion and rho masses ($m_\pi \simeq 0.57$ and $m_\rho \simeq 0.67$): $\frac{f_\pi}{m_\rho Z_A} = 0.226(11)$. The size of the statistical errors and the general flavour dependence allow to conclude that the inclusion of sea quarks tends to increase, especially for the $\rho$ case, the values of the (raw) decay constants.

By comparing the large and small volume results, one can conclude that the effect, which is visible only for moderately light quark masses, is stable against finite size corrections. Physical values of the light meson decay constants can be obtained only after extrapolating to zero lattice spacing and to the chiral limit, a scope beyond this work.

The extrapolation from negative flavour numbers appears smooth after matching a number of physical quantities corresponding to the renormalized parameters of the theory, agrees with the results obtained from direct fermion simulations and gives strong support for visible unquenching corrections for light meson decay constants.

| run label | $n_f$ | $(f_\rho Z_V)^{-1}$ | $a f_\pi / Z_A$ | $f_\pi / (m_\rho Z_A)$ |
|---|---|---|---|---|
| (a) | −4 | 0.322(8) | 0.166(4) | 0.235(6) |
| (b) | −2 | 0.324(7) | 0.168(4) | 0.238(6) |
| (c) | 0 | 0.328(8) | 0.166(4) | 0.234(6) |
| extrap. | 2 | 0.331(12) | | 0.234(9) |
| (d) | −4 | 0.373(9) | 0.160(5) | 0.243(8) |
| (e) | −2 | 0.394(6) | 0.166(5) | 0.251(8) |
| (f) | 0 | 0.418(9) | 0.173(5) | 0.263(8) |
| extrap. | 2 | 0.440(13) | | 0.271(13) |
| (g) | −4 | 0.371(7) | 0.157(3) | 0.234(5) |
| (h) | −2 | 0.405(8) | 0.168(5) | 0.255(8) |
| (i) | 0 | 0.424(9) | 0.172(6) | 0.264(10) |
| extrap. | 2 | 0.454(13) | | 0.283(13) |

Table 1: The values of the meson decay constants.

| run label | $n_f$ | Volume | $\beta$ | $K$ | $n_{conf}$ | $a m_\pi$ | $a m_\rho$ | $m_\pi^2 / m_\rho^2$ |
|---|---|---|---|---|---|---|---|---|
| (a) | −4 | $16^3 \times 32$ | 6.4 | 0.155 | 26 | 0.577(4) | 0.705(6) | 0.670(10) |
| (b) | −2 | $16^3 \times 32$ | 6.1 | 0.1557 | 20 | 0.586(3) | 0.706(7) | 0.689(11) |
| (c) | 0 | $16^3 \times 32$ | 5.767 | 0.1582 | 20 | 0.591(5) | 0.710(7) | 0.693(11) |
| (d) | −4 | $16^3 \times 32$ | 6.463 | 0.158 | 47 | 0.464(4) | 0.658(8) | 0.497(9) |
| (e) | −2 | $16^3 \times 32$ | 6.1 | 0.161 | 89 | 0.467(2) | 0.662(5) | 0.498(5) |
| (f) | 0 | $16^3 \times 32$ | 5.7 | 0.165 | 92 | 0.457(3) | 0.659(8) | 0.481(7) |
| (g) | −4 | $24^3 \times 32$ | 6.463 | 0.158 | 60 | 0.467(3) | 0.671(6) | 0.484(6) |
| (h) | −2 | $24^3 \times 32$ | 6.1 | 0.161 | 34 | 0.465(3) | 0.660(8) | 0.496(6) |
| (i) | 0 | $24^3 \times 32$ | 5.7 | 0.165 | 36 | 0.455(3) | 0.651(6) | 0.488(7) |

Table 2: The details of each run used to determine the decay constants values of Table 1. In runs (e) and (f) $f_\pi$ is measured only on 47 and 45 configurations respectively.

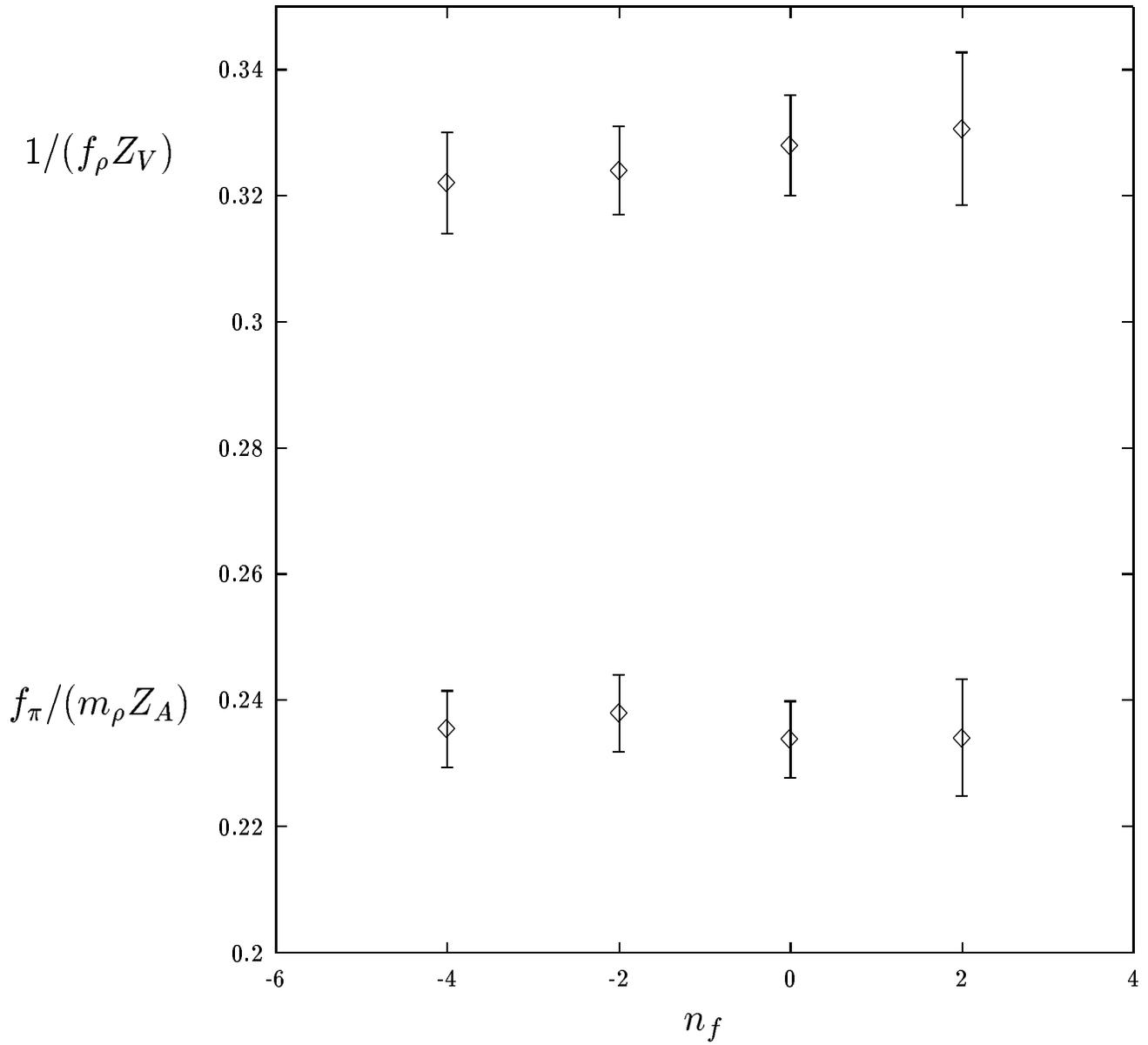

Figure 1: The flavour dependence of $f_\pi/(m_\rho Z_A)$ and $1/(f_\rho Z_V)$ decay constants for $(m_\pi/m_\rho)^2 = 0.7$

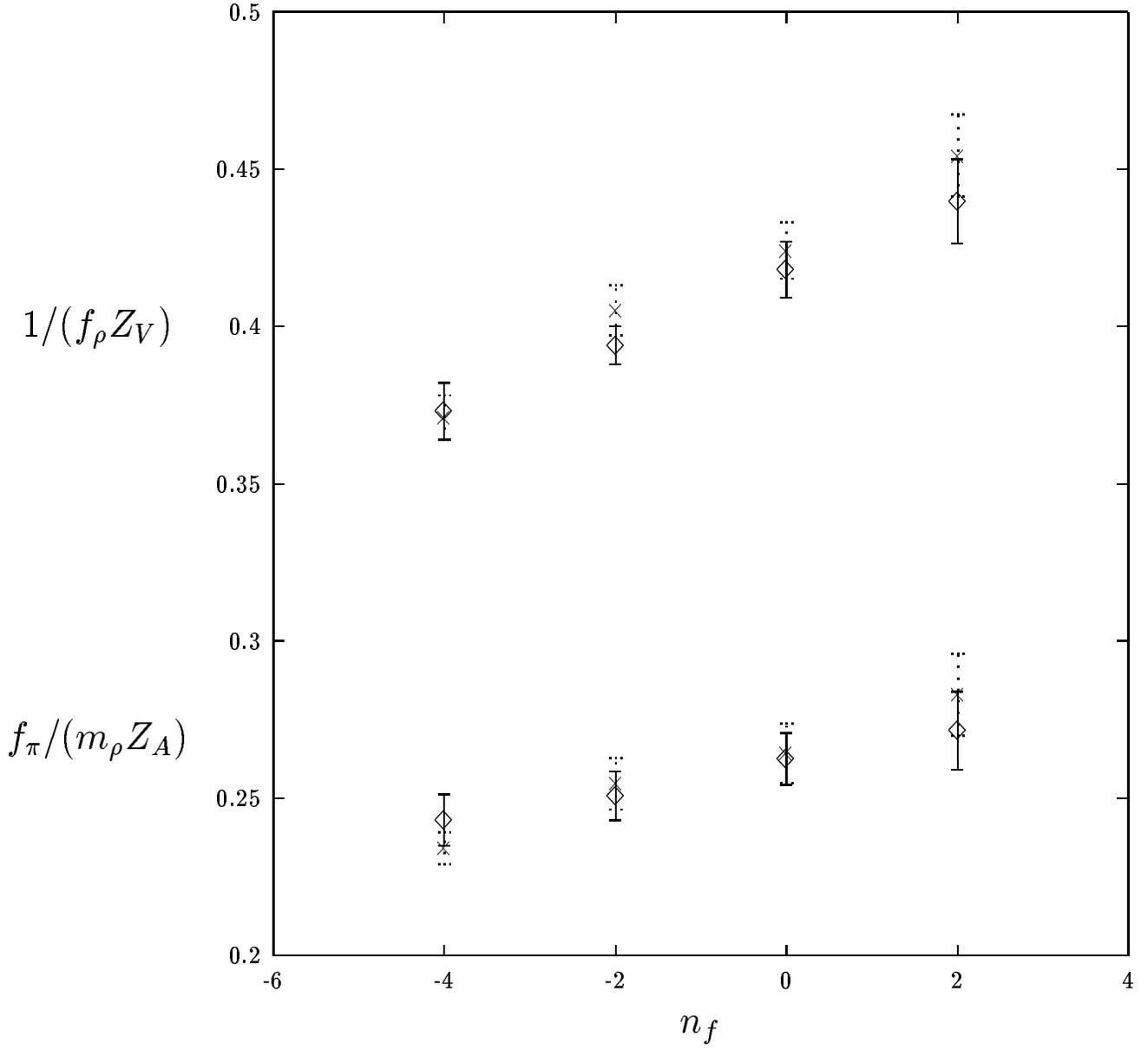

Figure 2: The flavour dependence of $f_\pi/(m_\rho Z_A)$ and $1/(f_\rho Z_V)$ decay constants for the $(m_\pi/m_\rho)^2 = 0.5$. Crosses refer to the $24^3 \times 32$ lattice and diamonds to the $16^3 \times 32$.

9